\documentclass[12pt]{article}
\usepackage{graphicx}
\usepackage{amssymb,amsmath}
\usepackage{epstopdf}
\usepackage{subfigure}
\usepackage{rotating}
\usepackage{color}
\usepackage{lscape,graphicx}
\usepackage{natbib}
\bibpunct{[}{]}{,}{n}{,}{,}

\def\blp{\bigg (}
\def\brp{\bigg )}
\def\la{\langle}
\def\ra{\rangle}
\def\j{\varphi}
\def\rv{{\mathbf{r}}}
\def\Rv{{\mathbf{R}}}
\def\D{\Delta}

\def\fv{{\mathbf{f}}}
\def\csch{{\mbox{csch}}}

\def\a{\alpha}
\def\g{\gamma}

\def\qv{{\mathbf{q}}}

\def\e{\epsilon}
\def\d{\delta}

\def\l{\lambda}
\def\erf{{\mbox{erf}}}
\def\fv{{\mathbf{f}}}

\def\N{{\cal{N}}}

\title{\bf\Large Stretching Homopolymers}
\begin{document}

\begin{center}
{\bf{\Large{Stretching Homopolymers}}}
\end{center}

\vspace{.5in}

\begin{center}
\begin{tabular}[t]{c@{\extracolsep{2em}}c}
  Greg Morrison$^{1,2}$, Changbong Hyeon$^3$, N. M. Toan$^2$, Bae-Yeun Ha$^4$, and D. Thirumalai$^{2,5}$\\
\end{tabular}
\smallskip
\begin{small}
\begin{tabular}{rl}
\parbox[t]{13cm}{
\begin{center}
$^1$  Department of Physics, University of Maryland at College Park, College Park, MD  20742\\
$^2$  Biophysics Program, IPST, University of Maryland at College Park, College Park, MD  20742\\
$^3$  Center for Theoretical Biological Physics, University of California at San Diego, La Jolla, CA  92093\\
$^4$  Department of Physics and Astronomy, University of Waterloo, Waterloo, Ontario, Canada N2L 3G1\\
$^5$  Department of Chemistry and Biochemsitry, University of Maryland at College Park, College Park, MD 20742
\end{center}
}\\
\end{tabular}
\end{small}
\bigskip
\end{center}

\baselineskip = 16pt

\begin{abstract}
Force induced stretching of polymers is important in a variety of contexts.  We have used theory and simulations to describe the response of homopolymers, with $N$ monomers, to force ($f$) in good and poor solvents.  In good solvents and for {{sufficiently large}} $N$ we show, in accord with scaling predictions, that the mean extension along the $f$ axis $\la Z\ra\sim f$ for small $f$, and $\la Z\ra\sim f^{\frac{2}{3}}$ (the Pincus regime) for intermediate values of $f$.  The theoretical predictions for $\la Z\ra$ as a function of $f$ are in excellent agreement with simulations for $N=100$ and 1600.  However, even with $N=1600$, the expected Pincus regime is not observed due to the the breakdown of the assumptions in the blob picture for finite $N$.  {{We predict the Pincus scaling in a good solvent will be observed for $N\gtrsim 10^5$}}.  The force-dependent structure factors for a polymer in a poor solvent show that there are a hierarchy of structures, depending on the nature of the solvent.  For a weakly hydrophobic polymer, various structures (ideal conformations, self-avoiding chains, globules, and rods) emerge on distinct length scales as $f$ is varied.  A strongly hydrophobic polymer remains globular as long as $f$ is less than a critical value $f_c$.  Above $f_c$, an abrupt first order transition to a rod-like structure occurs.  Our predictions can be tested using single molecule experiments.
\end{abstract}\newpage

\section*{I. Introduction}

Single molecule nanomanipulation methods have been used to measure the response of biological macromolecules to mechanical force.  Such measurements give direct estimates of the elasticity of DNA \cite{Buste}, RNA \cite{Liphardt}, proteins \cite{Gaub1,tibs}, and polysaccharides \cite{Gaub2}.  Although tension-induced stretching of RNA \cite{CBDT,HyeonStructure06,HyeonBJ07} and proteins \cite{HyeonStructure06,KlimovDT1} largely depends on the architecture of the folded conformations \cite{ KlimovDT2}, sequence effects \cite{Block} make it difficult to unambiguously interpret the measured force-extension curves (FECs) in terms of unfolding pathways.  In this context, stretching of homopolymers by force provides a potentially simpler case for which the FECs can be calculated.  

In a pioneering paper, Pincus \cite{Pincus} considered the strong stretching of homopolymers in a good solvent.  The strong stretching limit corresponds to a large enough force, $f$, such that $N^\nu a < \la Z\ra \ll N a$, where $\nu=3/5$ is the Flory exponent, $N$ is the number of monomers, $a$ is the size of a monomer, and $\la Z\ra=\la z_N-z_0\ra$ is the mean tension-induced end-to-end distance {{(we have assumed that $f$ is aligned with the $z$ axis)}}.  Pincus showed that the size of the stretched polymer should be determined by an interplay between the Flory radius $R_F=N^\nu a$ and the tensile screening length (or the blob size) \cite{Pincus}, $\xi_P=k_B T/f$.  When $f$ is small, then $x=R_F/\xi_P\ll 1$, while in the opposite limit, $x\gg 1$.  The scaling assumption is that for arbitrary $f$, the average end-to-end distance 
can be written as
\begin{eqnarray}
\la Z\ra=R_F\Phi(R_F/\xi_P)
\end{eqnarray}
With this assumption, one can anticipate three regimes in the FEC.

($i$)  For small $f$, we expect a linear increase in the extension of the chain with $\fv$ in the $z$-direction.  At low forces, $\Phi(x)\approx x$, and hence $\la Z\ra\propto R_F^2\times (a\beta f)$.    
($ii$) In the strongly stretched limit, which arises for intermediate forces, the value of $\la Z\ra$ can be obtained by dividing the chain into a sequence of aligned tensile blobs (along the force axis) {whose size is $\xi_P\sim (\beta f)^{-1}$} \cite{deGennesBook}.   The monomers contained within each blob behave as an unperturbed self-avoiding walk.  In this case $\xi_P=(\beta f)^{-1}\sim N_b^{\nu}$, with $N_b$ the number of monomers in a blob.  The linear extension of the chain is then given by $\la Z\ra\sim \xi_P\times N/N_b\sim N(\beta f)^{\frac{1}{\nu}-1}\sim N(\beta f)^{\frac{2}{3}}$.  We will refer to this intermediate scaling regime as the Pincus regime.  It should be stressed that this argument is valid only if $N\gg(\xi_P/a)^{\frac{1}{\nu}}\gg 1$, which may not be satisfied for a stiff polymer, or a flexible polymer with small $N$ (see below).  ($iii$)  For extremely large forces {{(beyond the strong stretching regime)}}, we expect the excluded volume to become irrelevant, as the bonds between monomers become fully aligned with the $z$-axis, and no monomer interacts with any other monomer.  The FEC in this regime will be model-dependent, with $\la Z\ra\approx N a^2\beta f/3$ for an extensible chain, and $\la Z\ra\approx N a$ for a inextensible chain.   We will refer to this behavior as the non-universal regime.

The Pincus scaling description of the stretching of homopolymers is well known.  However, as far as we are aware, a microscopic derivation of the FEC anticipated by Pincus has not been provided.  More importantly, it is unclear how the FEC of polymers with finite $N$ compares with the predictions of the scaling theory.  In other words, for finite values of $N$ (on the order of 1000), how pronounced is the Pincus regime?  In this paper, we develop a self-consistent, variational theory based on the Edwards-Singh method \cite{ES} to determine the average extension of a homopolymer in a good solvent.  The theory gives excellent agreement with simulations.  Surprisingly, neither the theoretical predictions nor simulations display the Pincus regime for $N=100$ or $N=1600$.  We show that this is due to a finite-size effect, and show that the Pincus regime emerges only for $N\gtrsim 10^5$.  Only when $N$ is large is the concept of the tensile blob (with $\xi_P\sim a N_b^\nu$) satisfied, where $N\gg N_b\gg 1$.  We also show using theory and simulations that the tension induced stretching of homopolymers in a poor solvent exhibits a first order transition between an ensemble of collapsed states and rod-like conformations.  The nature of the transition is dependent on how poor the solvent is, which is measured in terms of the relative attraction between the monomers.  The theoretical predictions for the poor solvent case are only in qualitative agreement with the simulations.  Simulations of a polymer in a poor solvent show that tension-induced transitions occur via a hierarchy of structures, depending on the solvent quality.  Force-dependent structure factors show that, for a weakly hydrophobic polymer, the transition to the stretched state occurs through a variety of structures, depending on the length scale (or the magnitude of the wave vector, $q$).  For a strongly hydrophobic chain, the globule to rod transition occurs by a first order transition when $f$ exceeds a critical value. \\

\section*{II. Polymers Under Tension in a Good Solvent}

\subsection*{Theory}
{\it {Extensible Polymer: } }The Hamiltonian for a self-avoiding polymer chain under tension is taken to be
\begin{eqnarray}
\beta H_0&=&\frac{3}{2a^2}\int_0^N ds\ \dot\rv^2(s)-\beta f\int_0^N ds\ \dot z(s)+\D_2,\label{HG}
\end{eqnarray}
where {{$f$ is aligned with the $z$-axis,}} $\beta=1/k_B T$, and
\begin{eqnarray}
\D_2&=&\frac{v_0}{2}\int_0^N ds\int_0^Nds' \ \d[\rv(s)-\rv(s')],
\end{eqnarray}
with $v_0$ the strength of the self-avoiding interaction, with $v_0>0$ in a good solvent.  To compute the force-extension curves (FECs) {{and compare them to simulations}}, we use a self-consistent variational method, {{originally proposed by}} Edwards and Singh \cite{ES}.  Following the convention in single molecule experiments, we use FEC for the extension changes upon application of force.  However, throughout the paper, we will derive and plot the extension $\la Z\ra$ as a function of $f$.  A reference Hamiltonian
\begin{eqnarray}
\beta H_1&=&\frac{3}{2a^2\l^2}\int_0^N ds\ \dot\rv^2(s)-\beta f\int_0^N ds\ \dot z(s)\label{HV},
\end{eqnarray}
is chosen and the parameter $\l$ is determined self-consistently.  Because we are interested in calculating the FECs, the {{relevant quantity}} is the dependence of $\la Z(f)\ra_0=\la z_N-z_0\ra=\int_0^N ds\,\la\dot z(s)\ra_0$ on $f$, where $\la\cdots\ra_0$ indicates the Boltzmann-weighted average with respect to $\beta H_0$.  In Appendix A, we also consider the square of the transverse fluctuations using the Edwards-Singh method.  Because it is not possible to compute the exact average $\la Z\ra_0$, we calculate the difference between $\la Z\ra_0$ and $\la Z\ra_1$ (where $\la \cdots\ra_1$ is the average with respect to $\beta H_1$), assuming that $\Delta_1+\Delta_2$ is small, with 
\begin{eqnarray}
\D_1&=&\frac{3}{2a^2}\blp 1-\frac{1}{\l^2}\brp\int_0^N ds\ \dot\rv^2(s).
\end{eqnarray}
To first order in $\D_1+\D_2$, we obtain
\begin{eqnarray}
\la Z\ra_0-\la Z\ra_1=\la Z(\D_1+\D_2)\ra_1-\la Z\ra_1\la\D_1+\D_2\ra_1.
\end{eqnarray}
A self-consistent equation for $\l$ is obtained by insisting that $\la Z\ra_0\approx\la Z\ra_1$, which leads to the condition
\begin{eqnarray}
\la Z(\D_1+\D_2)\ra_1=\la Z\ra_1\la\D_1+\D_2\ra_1.\label{sc}
\end{eqnarray}
Throughout this work, we compute averages with respect to $H_1$, so the subscripts on $\la\cdots\ra$ will be dropped.  The terms involving $\D_1$, $\D_2$ are easily calculated using 
{\begin{equation}
\la Z\D_1\ra-\la Z\ra\la\D_1\ra=\frac{1}{2}\l(\l^2-1)\ \frac{\partial\la Z\ra}{\partial\l}=\frac{\lambda^2(\lambda^2-1)Na^2}{6}\beta f
\label{Del1}
\end{equation}
and 
\begin{equation}
\la Z \D_2\ra-\la Z\ra\la\D_2\ra=\frac{v_0}{2}\int_0^N ds\int_0^N ds'\frac{\partial}{\partial(\beta f)}\la\d [\rv(s)-\rv(s')]\ra
\label{Del2}
\end{equation}
with $\la\delta[\rv(s)-\rv(s')]\ra=({3}/{2\pi a^2\l^2|s-s'|})^{\frac{3}{2}}\exp{\left(-|s-s'|\l^2a^2\beta^2 f^2/6\right)}$ (the details of the calculations are given in Appendix A).  Using Eqs. (\ref{sc}), (\ref{Del1}), and (\ref{Del2}),} the self-consistent equation for $\l$ becomes
\begin{eqnarray}
\l^2-1&=&\frac{v\sqrt{N}}{\l^3}\int_\d^1 du\frac{1-u}{\sqrt{u}}\ e^{-Nu\l^2\j^2/6}\nonumber\\
&=&\frac{6v}{\l^5\j^2\sqrt{N}}\bigg\{ e^{-N\l^2\j^2/6}-\sqrt{\d}\,e^{-\d N\l^2\j^2/6}\nonumber\\
&&\qquad\qquad+\frac{\j}{\l^3}\sqrt{\frac{N\pi}{6}}\blp 1-\frac{3}{N\l^2\j^2}\brp\bigg[\erf\blp\l\j\sqrt{\frac{N}{6}}\brp-\erf\blp\l\j\sqrt{\frac{\d N}{6}}\brp\bigg]\bigg\},\label{GaussianConsistent}
\end{eqnarray}
where we have defined the dimensionless excluded volume parameter $v=(3/2\pi)^{\frac{3}{2}}\ v_0/a^3$, the dimensionless force $\j=a\beta f$, and where $\erf(x)$ is the error function.   We have also included a cutoff, $\d$, in the integral over $u$ (with $u=|s-s'|/N$), to account for the finite separation between the monomers, {which is neglected in the continuum representation of the Hamiltonian in {Eq.} (\ref{HV})}.  We expect $\d\sim \l/N$, since the discrete monomers are separated by a distance $|\rv_{i+1}-\rv_i|\approx \l a$ on an average in the reference Hamiltonian, $\beta H_1$.  The cutoff is only {{imposed in}} theories that have a self-energy divergence \cite{ES,Cordeiro,Cylinder,HaThirum}, and is {{generally}} not required if there is no divergence, as is the case here.  However, we will see that this cutoff is essential in order to reproduce the FECs obtained in simulations.  Given a solution $\l$ to the SCE, the linear end-to-end distance is given by $ \la Z\ra=Na^2\l^2\beta f/3$.

It is not difficult to show that, as $f\to 0$, {a solution to {Eq.} (\ref{sc}) is} $\l\approx\l_0 \propto(v^2 N)^{\frac{1}{10}}$, giving the expected linear regime, $\la Z\ra\sim N^{\frac{6}{5}}v^{\frac{2}{5}}{a}\times(a\beta f)$.  We immediately see that this gives the correct scaling with $N$ and $v$ for low forces, with $\la Z\ra\approx \la\Rv^2\ra_{f=0}\times (\beta f)/3$.  We also note that, if we set $\delta =0$,  we exactly recover {{(in our notation)}} the original, tension-free self-consistent equation for a self-avoiding chain, $\l^2-1=\sqrt{6N/\pi^3}\, v_0a^3/\l^3$ developed by Edwards and Singh \cite{ES}.

For intermediate $f$, we can obtain the correct Pincus scaling for large $N$.  If we assume $\l\approx \l_0$, we find $N\l_0^2\j^2\gg 1$ when $\j\approx a\beta f_T\sim N^{-\frac{3}{5}} v^{-\frac{1}{5}}$, defining the transition force $f_T$ into the strongly stretched Pincus regime.  {{For $f\ge f_T$}}, we can neglect terms on the order $N^{-1}$ {{and}} $\exp(-N\l^2\j^2/6)$ {{for large $N$}}, and set $\erf(\l\j\sqrt{N/6})\approx 1$.  This gives the approximate SCE  
\begin{eqnarray}
\l^2-1\approx \frac{v\sqrt{6\pi}}{\l^4\j}\bigg[ 1-\erf\blp\l\j\sqrt{\frac{N\d}{6}}\brp\bigg] +O(N^{-1}).
\end{eqnarray}
With $\d\sim \l/N$, we see that we can neglect the error function in this regime as well if $a\beta f_T\sim \l_0^{-\frac{3}{2}}v^{\frac{1}{10}}N^{-\frac{9}{20}}\ll 1$.  If $N$ is sufficiently large to satisfy this requirement, the SCE becomes $\l^2-1\approx v\sqrt{6\pi}/\l^4\j+O(v N^{-\frac{1}{4}})$.  We thus find the approximate solution {{in the Pincus regime}} $\l\approx \l_P\propto (v/\j)^{\frac{1}{6}}$.  For large $N$ and intermediate forces, we find $\la Z\ra\propto N v^{\frac{1}{3}}f^{\frac{2}{3}}$, as is expected \cite{Pincus}.  Note that neglecting terms of order $v N^{-\frac{1}{4}}$ may be valid only for extremely large $N$ (on the order of $N\sim 10^5$).  Thus, the onset of the non-linear scaling regime depends on both $v$ and $N$, as was anticipated by Pincus.  

For sufficiently large $\j$, we can neglect terms of order $\j^{-1}$ in Eq. (\ref{GaussianConsistent}), to find {{an extended or rod-like solution}} $\l\approx\l_{E}= 1$.  This root gives $\la Z\ra\approx Na^2\beta f/3$, identical to the non-interacting average for an extensible chain.  This is not surprising; as the tension becomes large, the excluded volume interaction is not relevant.  We also note that, in this regime, the chain will become greatly overextended.  As was shown by Pincus, the extension beyond the non-linear regime is non-universal and depends on the precise model used for the homopolymer \cite{Pincus}.

{\it{Inextensible Polymer:  }}Because the extensible polymer can overstretch for large forces, which may not occur for real polymers that are linked by covalent bonds with high spring constants, we develop a theory for an approximately inextensible model.  We were also motivated to consider the inextensible model because the Monte Carlo simulations for $N=1600$ (see Appendix B) were performed for a model in which the distance between successive beads is precisely $a$.  We begin with the discrete, non-interacting, spring-like Hamiltonian
\begin{eqnarray}
H[\{\rv_n\}]=\frac{k}{2a^2}\sum_n (|\D\rv_n|-a)^2-\beta f\sum_n\D z_n.\label{SpringDiscrete}
\end{eqnarray}
The average end-to-end distance, as well as fluctuations in the $x$ and $z$ directions are easily computed using this Hamiltonian.  Defining $X=x_N-x_0$, we find
\begin{eqnarray}
\frac{\la Z\ra}{Na}&=&\frac{1}{\N}\int_0^\infty dx\ x^2 e^{-k(x-1)^2/2}\cosh(\j x)-\frac{1}{\j}\label{avz},\\
\frac{\la Z^2\ra-\la Z\ra ^2}{Na^2}&=&\frac{1}{\N}\int_0^\infty dx\ x^3 e^{-k(x-1)^2/2}\sinh(\j x) -\blp\frac{\la Z\ra}{Na}\brp^2-\frac{2}{\j}\blp\frac{\la Z\ra}{Na}\brp,\label{zfluct}\\
\frac{\la X^2\ra}{Na^2}&=&\frac{1}{\j}\blp\frac{\la Z\ra}{Na}\brp,\label{xfluct}\end{eqnarray}
{where $\N=\int_0^\infty dx\ x e^{-k(x-1)^2/2}\sinh(\j x)$, and} we have used $\la Z^2\ra=N\la z_n^2\ra+N(N-1)\la z_n\ra^2$ in Eq. (\ref{zfluct}).  We approximate the Hamiltonian in Eq. (\ref{SpringDiscrete}) with a continuous chain using an Inextensible Gaussian Hamiltonian (IGH) \cite{Benoit, Hatfield}
\begin{eqnarray}
H_I[\rv(s)]=\frac{3}{2a^2}\int_0^N ds\blp\frac{\dot x^2(s)+y^2(s)}{\a_1^2(k,\j)}+\frac{\dot z^2(s)}{\a_3^2(k,\j)}\brp-\beta g(k,\j)\int_0^N ds \dot z(s)\label{InextensH}
\end{eqnarray}
where $\a_1$ and $\a_3$ are the effective spring constants in the longitudinal and transverse directions, respectively, and $g$ is an effective tension.  The spring constants $\a_1$ and $\a_3$, and the effective tension $g$, are functions of $k$ and $\j$.  Using {{the IGH}}, we find
\begin{eqnarray}
\la Z\ra&=&N a \a_3^2 \beta g/3,\nonumber\\
\la Z^2\ra-\la Z\ra^2&=&Na^2\a_3^2/3,\nonumber\\
\la X^2\ra&=&Na^2\a_1^2/3.
\label{modav}
\end{eqnarray}
Equating the averages in {Eq.} (\ref{modav}) with those in {Eqs.} (\ref{avz})-(\ref{xfluct}) {{explicitly}} gives the desired IGH {{in terms of $k$ and $f$}}.  The full expression for the $\a_i$'s and $g$ are quite lengthy for general $k$ and $f$, and we omit them here.  Note that, with an insertion of $\d(x-1)$ into all integrals in {Eqs. (\ref{avz})-(\ref{xfluct})}, or equivalently, in the limit as $k\to\infty$, we recover the Freely Jointed Chain (FJC) averages.  In the FJC limit, the expressions for the $\a_i$'s and $g$ are quite simple, and we find
\begin{eqnarray}
\a_1^2&=&\frac{3}{\j^2}\blp\j\coth(\j)-1\brp,\nonumber\\
\a_3&=&\frac{3}{\j^2}\blp 1-\j^2\csch^2(\j)\brp,\nonumber\\
\a\beta g&=&\j\,\frac{\a_1^2}{\a_3^2}.
\end{eqnarray}
These spring constants, $\a_i(k\to\infty,f)$, were derived by Hatfield and Quake using a different method \cite{Hatfield}.  

We note that this approximate FJC Hamiltonian gives the simple Gaussian behavior for $\j\to 0$, whereas in the limit of $\j\to\infty$, we can easily show that the distributions give the expected form of $P(X)=\d(X)$, $P(Y)=\d(Y)$, and $P(Z)=\d(Z-Na)$, {{with $X=x_N-x_0$, and similarly for $Y$.}}  We therefore expect that the IGH to be an excellent approximation for an inextensible chain in  the limits of small and large $f$, with possible deviations from the correct distribution for intermediate $f$.  Because of the more complicated form of the Hamiltonian in Eq. ($\ref{InextensH}$), exact analytic work is difficult in the inextensible case.  We can, however, generate a self-consistent equation using Eq. (\ref{sc}) to determine the FEC of a self-avoiding inextensible chain in a manner similar to the extensible case.  Using the reference Hamiltonian 
\begin{eqnarray}
H_r=\frac{3}{2a^2\l^2}\int_0^N ds\blp\frac{\dot x^2(s)+y^2(s)}{\a_1^2}+\frac{\dot z^2(s)}{\a_3^2}\brp-\beta g\int_0^N ds\, \dot z(s),\label{InexRef}
\end{eqnarray}
and defining
\begin{eqnarray}
\D_1^{(IGH)}=\frac{3}{2a^2}\blp1-\frac{1}{\l^2}\brp\int_0^N ds\blp\frac{\dot x^2(s)+y^2(s)}{\a_1^2}+\frac{\dot z^2(s)}{\a_3^2}\brp,
\end{eqnarray}
we can, to first order in $\D_1^{(IGH)}+\D_2$, develop the self-consistent equation $\la Z(\D_1^{(IGH)}+\D_2)\ra=\la Z\ra\,\la\D_1^{(IGH)}+\D_2\ra$, similar to Eq. ($\ref{sc})$.  
The form of the inextensible SCE is similar to that of ($\ref{GaussianConsistent}$), with
\begin{eqnarray}
\l^2-1=\frac{v\sqrt{N}}{\l^3\a_1^2\a_3}\int_\d^1 du\ \frac{1-u}{\sqrt{u}} e^{-N\l^2\a_3^2\g^2 u/6},\label{modsc}
\end{eqnarray}
with $\g=a\beta g$ the dimensionless effective tension.  It is possible, albeit complicated, to show that the solution to Eq. (\ref{modsc}), with $k\gg 1$, will be divided into approximately the same scaling regions as we found in the extensible case.  The solutions to the inextensible SCE, determined using Eq. (\ref{modsc}), are similar to the extensible roots from Eq. (\ref{sc}), with significant differences in the two models occurring only for $\j\gtrsim 1$.  Again, the expected Pincus scaling of $\la Z\ra\sim f^{\frac{2}{3}}$ emerges only for very large $N$.  Thus, both for the extensible chain and the IGH with excluded volume interactions, the linear behavior and the Pincus regime are obtained.  The behavior {{of the FEC}} in the limit of very large force is clearly model dependent, {{as predicted by Pincus \cite{Pincus}}}.   The theoretical predictions for the IGH with excluded volume interactions are validated by explicit comparison to Monte Carlo simulations {{(see below)}}.\\

\subsection*{Simulations}
{\it{Extensible Polymer:  }}In order to determine if the theory accurately predicts the effect of excluded volume on a self-avoiding polymer under tension, we have performed Langevin simulations with $N=100$ at various stretching forces. To calculate the equilibrium FEC of a self-avoiding polymer, we performed low friction Langevin dynamics simulations using the Hamiltonian
\begin{equation}
\beta H=\frac{3}{2a^2}\sum_{i=1}^{N-1}(|\rv_{i}-\rv_j|^2-a^2)+\sum_{i=1}^{N-2}\sum_{j=i+2}^N\varepsilon \left(\frac{a}{|\rv_{i+1}-\rv_i|}\right)^{12}-\beta f(z_N-z_1)\label{GoodSolvHam},
\end{equation}
with $a=1$, $\varepsilon=100$, and $N=100$.  We set $k_BT=1/\beta=1$ in the simulations. The first term in Eq. (\ref{GoodSolvHam}) describes the chain connectivity in the extensible form that, in the continuum limit, becomes $3/2a^2\int_0^N ds\, \dot\rv^2(s)$.  We model the excluded volume interactions between the monomers using a $r^{-12}$ repulsion term (the second term in Eq. (\ref{GoodSolvHam})). Because of the large $\varepsilon$ value, the summation does not include neighboring monomers ($i$ and $i+1$) to avoid excessive repulsive forces. 
The last term in Eq. (\ref{GoodSolvHam}) denotes the potential due to tension acting on the ends of the polymer.  Thus, this model can be viewed as the discrete representation of the Hamiltonian in Eq. (\ref{HG}).  

\begin{figure}
\begin{center}
\includegraphics[width=0.5\textwidth]{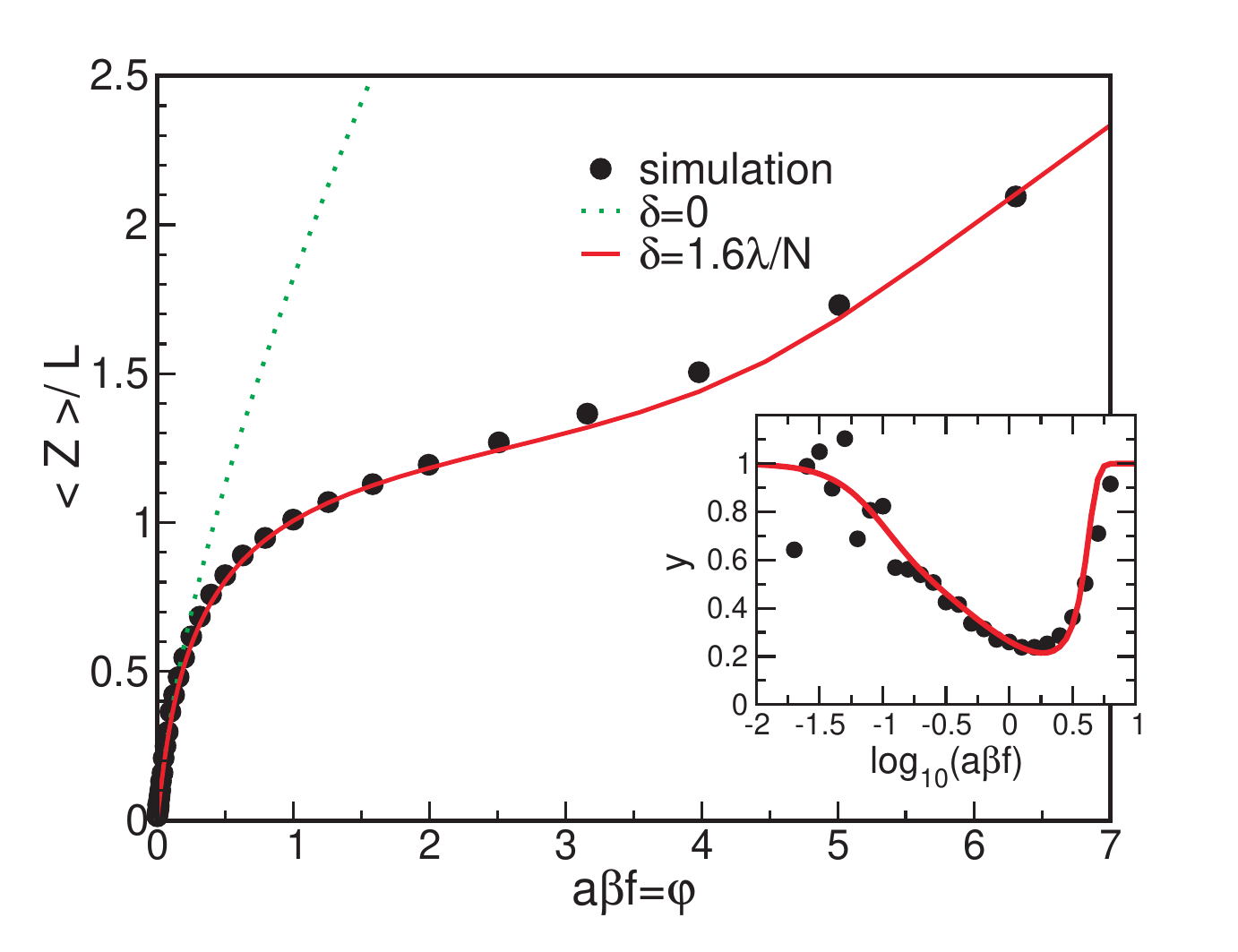}
\caption{$\la Z\ra$ as a function of $\j$ for varying $\j$.  The dots are the simulation results with $N=100$.   The linear regime corresponds to the dimensionless excluded volume parameter $v\approx 58.6$.  The best visual fit (solid line) is obtained with  $\d=1.6\l/N$.  Also shown are the fits with $\d=0$ (dotted line).  The inset compares the theoretical predictions (solid line) and the simulations results (dots) for the effective scaling exponent $y$.}
\end{center}
\end{figure}

The Langevin equations for each monomer are integrated in the low friction limit, which has been shown to accelerate the sampling rate of the conformational space of the polymer \cite{HoneycuttBP92}.  The equations of motion are
\begin{equation}
m\ddot{{\rv}_i}=-\zeta\dot{{\rv}_i}-\frac{\partial H}{\partial{\rv}_i}+\vec{\Gamma}_i \ ,
\end{equation}
where $m$ is the mass of the monomer, $\zeta$ is the friction coefficient, $-{\partial H}/{\partial{\rv}_i}$ is the conformation force arising from Eq. (\ref{GoodSolvHam}), and $\vec{\Gamma}_i$ is a random force that satisfies the fluctuation-dissipation theorem, $\langle\vec{\Gamma}_i(t)\cdot\vec{\Gamma}_j(t')\rangle={6\zeta k_BT}/{h}\ \delta(t-t')\delta_{ij}$, where the integration time ($h$) is discretized.  The natural time is $\tau_L=(ma^2/\varepsilon_h)^{1/2}$.  We chose $\zeta=0.05\tau_L^{-1}$ and $h=0.002\tau_L$.  To begin the simulations, we generate 200 initial random polymer conformations, and thermally equilibrate those structures for $5\times 10^6$ $h$ with $f=0$. Subsequently, a constant force is applied in the $z$-direction to one end of each polymer, with the other end held in a fixed position. The force exerted is increased as $f_j=10^{-3+0.1 j}$ $k_BT/a$ with $j=1,2,\ldots, 39$. The integer $j$ is increased every $5\times 10^6$ $h$.  {{For each force step}}, we neglect the first $2\times 10^6$ steps to ensure that the chain has equilibrated at $f_j$, and collect the statistics of polymer conformations every $10^4$ integration time steps for the remaining time steps.

In order to compare our theory to simulations, we need two {{fitting}} parameters, $v_0$ and $\d$.  We determine $v_0$ by fitting the {{simulated FEC in the linear, low force regime}}, and obtain $\d$ by a global fit of the theoretical predictions of the FEC to the results of the simulation.   The scaling laws for the extension as a function of force can not be accurately determined by simply fitting a linear \cite{Toan1,Toan2,Matthai} or log-log \cite{PRA} plot of the FEC.  Such fits implicitly assume that there exists a well-defined scaling regime, where $\la Z\ra\sim f^y$ with $y$ constant.  In order to determine the various scaling regimes of the FEC without imposing such an assumption, we will define the force-dependent effective scaling exponent $y$ such that
\begin{eqnarray}
y=\frac{\partial \log(\,\la Z\ra\,)}{\partial \log(\j)}.\label{ExpDef}
\end{eqnarray}
In Figure 1, we show the best fit of the theory compared to the simulations for the polymer in a good solvent ($v_0>0$).  With the choice of $v\approx 58.6$ and $\d=1.6\l/N$, the theoretical predictions agree well with the simulation data.  We note that this gives $v_0\approx 178 a^3$, significantly larger than the hard-core second virial coefficient of $v_0=4\pi a^3/3$.  It is known from the Edwards-Singh calculation \cite{ES} (with $f=0$) that if higher order terms are included in deriving the self-consistent equation (Eq. ($\ref{sc}$)), they merely renormalize $v$ without altering the scaling behavior.  A similar behavior is expected when $f\ne 0$.  As a result of the renormalization of $v$, we find that the extracted value of $v$ from simulations is larger than the naive value calculated from the second virial coefficient.  

We see that the theoretical predictions depend very strongly on the choice of cutoff, {{with the $\d=0$ theoretical FEC showing very poor agreement with the simulated data for $a\beta f\gtrsim 0.1$}}.  This is somewhat surprising, as a cutoff in the continuum limit approximation generally is used only to avoid self-energy divergences in the theory \cite{ES,Cordeiro,Cylinder,HaThirum}, which are not present here.  We also note that neither the theory nor the simulation predicts a Pincus-like scaling of $y\approx 2/3$, because the notion of the unperturbed tensile blob is not applicable for $N=100$ (see below).  

\begin{figure}
\begin{center}
\includegraphics[width=0.45\textwidth]{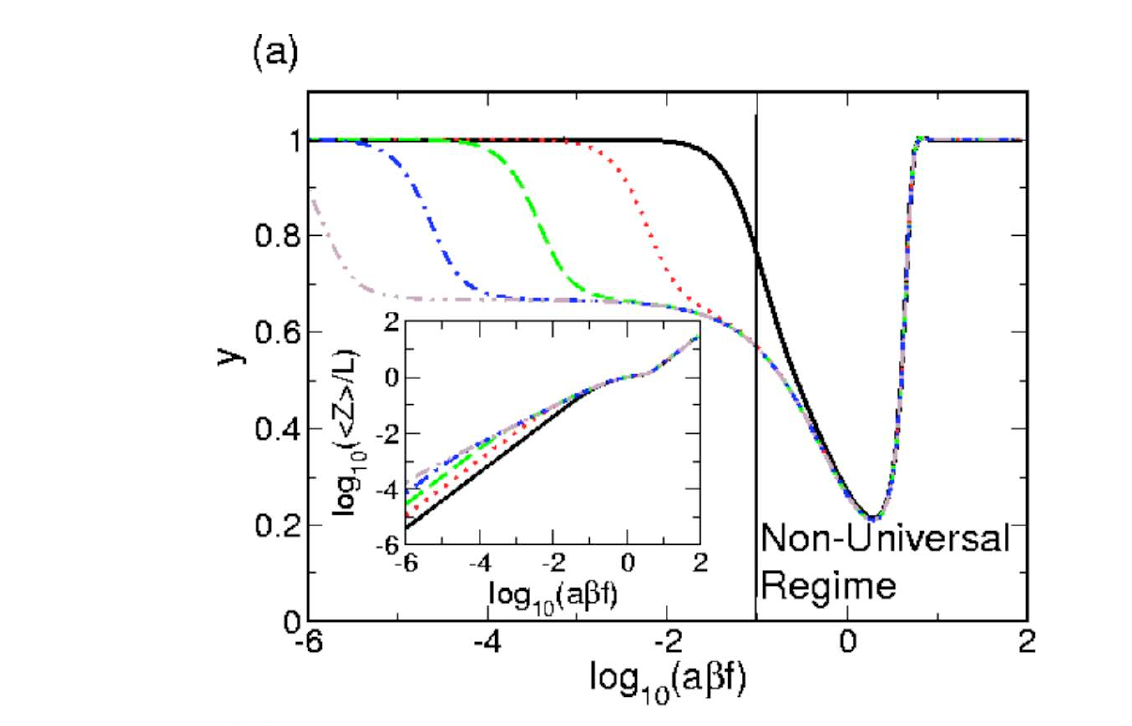}\qquad \includegraphics[width=0.45\textwidth]{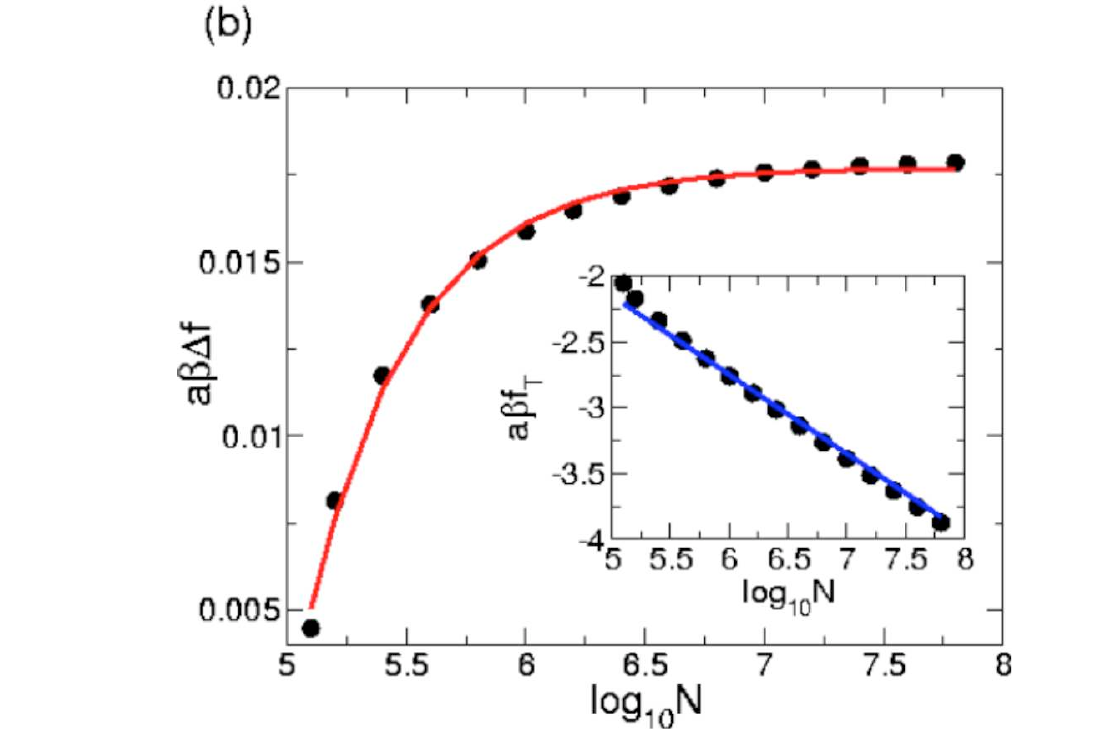}
\caption{(a) The effective scaling exponent $y$ for $N=10^2$ (\textemdash), $N=10^4$ ($\cdots$), $N=10^6$ ($- -$), $N=10^8$ ($-\cdot -$), and $N=10^{10}$ ($-\cdot\cdot -$), all with $v=58.6$ and $\d=1.6\l/N$ obtained theoretically.  The inset shows the log-log plot of the extension vs.  force, for the same parameters. (b) The width of the Pincus regime $\D f$ as a function of $N$ for $\epsilon = 0.05$.  The inset shows the initial Pincus transition force $f_T$ as a function of $N$.  Also shown is the predicted $N^{-\frac{3}{5}}$ scaling.}
\end{center}
\end{figure}

In order to asses the conditions under which the Pincus regime {{can be}} obtained, we plot the theoretical effective scaling exponent $y(\j$) for increasing $N$ in Fig. 2 (a).  While there is no clear Pincus regime for N=100, the expected 2/3 scaling emerges for larger $N$.  Variation in $v$ (i.e. changing the interaction strength of the excluded volume) only effects the depth of the trough (see Fig. 2(a)) in the final transition (data not shown), so adjusting $v$ can not yield the expected Pincus scaling for smaller $N$.  Figure 2 also shows that a very large $N\sim 10^6$ is required in order to see the 2/3 scaling over a large force range. For small values of $N$, the inequality $N\gg(\xi_P/a)^{\frac{1}{\nu}}\gg 1$ required to observe the Pincus scaling is not satisfied.  The width $\D f$ over which the strong stretching is observed can be computed using the self-consistent theory.  If we define the Pincus regime such that $\partial y/\partial \j\le\e$ (with $y$ defined in Eq. (\ref{ExpDef})) for some tolerance $\e$, we can numerically determine the dependence of the width of the Pincus regime with respect to $N$.  The width of the Pincus regime, $\D f$, is shown in Fig. 2 (b), along with a fit $\D f\approx 0.018-1600 N^{-1}$.  In the inset, we show the transition force into the Pincus regime, $f_T$, along with the expected scaling of $N^{-\frac{3}{5}}$.  We can extrapolate that the minimum number of monomers, $N_{min}$, for a self-avoiding polymer to show that the Pincus regime emerges only when $N_{min}\approx 9\times 10^4$ for $\e=0.05$.  Larger values of $N$ are required for the Pincus scaling to continue over an observable interval of $f$.  This finite size effect is remarkable, because when $f=0$ the exponent ($\nu\approx 0.6$) can be accurately obtained with $N<100$ \cite{Kalos}.  Because $N_{min}$ is too large for accurate simulations, it is not possible to explicitly demonstrate the nonlinear scaling {\it{in silico}}.  In principle, single molecule AFM or optical tweezer experiments can be used to confirm the predictions.\\

{\it{Inextensible Polymer:  }}In order to test our inextensible theory, we determine the best fit to a Monte Carlo simulation of a thick chain  \cite{Toan1,Toan2} with $N=1600$.  The thick chain is an inextensible, hard-core excluded volume model, with a configuration rejected if a triplet of monomers lie within a circle of radius $a$ (see Appendix B for details).  {{Our variational Hamiltonian in Eq. (\ref{InexRef}) is generated using the spring constant $k=10^4$ in Eq. (\ref{InextensH}).  In Fig. 3, we compare the FEC and effective scaling exponent (Eq. (\ref{ExpDef})) for the simulations and the inextensible theory, in Eq. (\ref{modsc}).}}  The FEC obtained using Monte Carlo simulations is in very good agreement  with the theoretical predictions (Fig. 3(a)).  We find $v\approx 15.7$ gives a good fit for the simulation data for low forces, {{and again $\delta=1.6\l/N$ gives a good global fit to the simulated data.}}

\begin{figure}
\begin{center}
\includegraphics[width=0.9\textwidth]{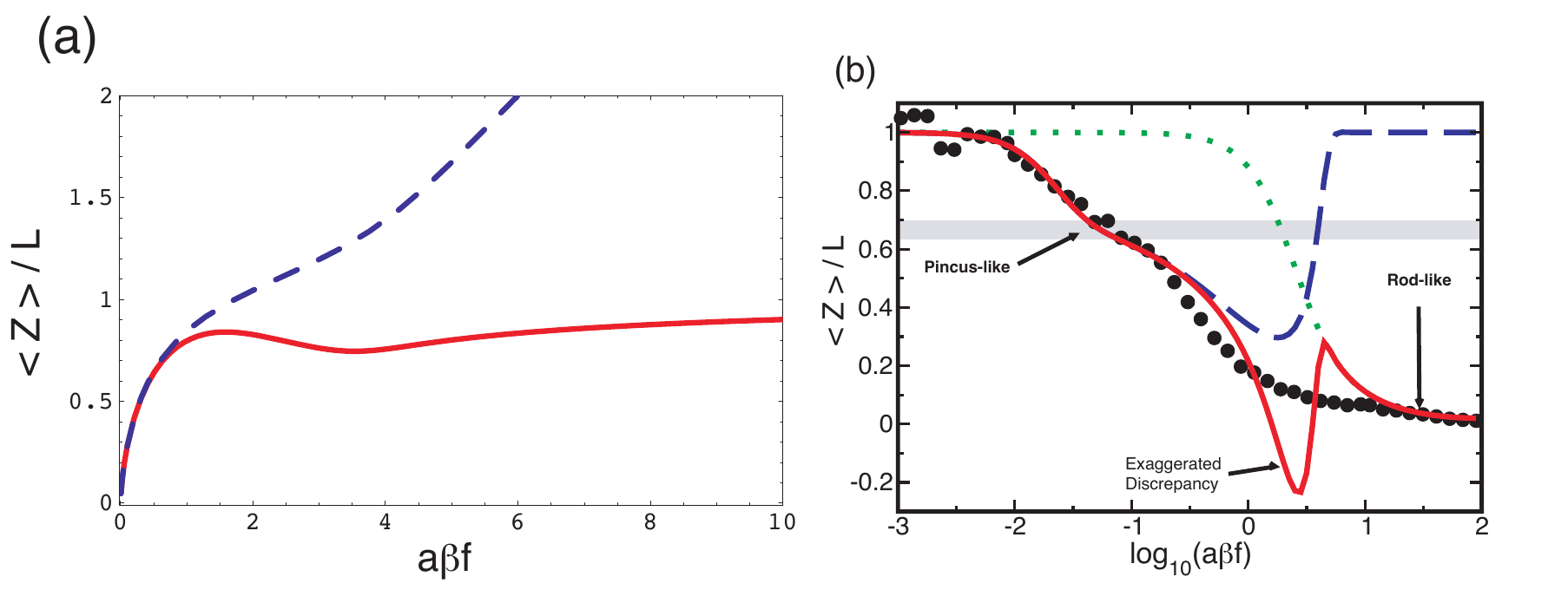}
\caption{(a)  Force extension curve for an inextensible chain with $N=1600$.  Shown are the simulation data (dots), along with the best fit for the IGH (solid line) and extensible Hamiltonian (dashed line), with $v=15.7$ and $\d=1.6\l/N$.  (b):  Effective scaling exponent $y$ for the inextensible FJC.  The solid line shows the theoretical exponents for the IGH and the dashed lines correspond to  the extensible Hamiltonian.  Also shown is the non-interacting FJC exponent (dotted line).}
\end{center}
\end{figure}
In Fig 3(b), we see that there is a deflection in $y\approx 2/3$ at $\j\approx 0.1$, corresponding to a Pincus-like regime observed in the simulations, and predicted by the theory based on Eq. (\ref{modsc}).  {{Such a deflection near $y=2/3$ is predicted by the theory for both the extensible and inextensible Hamiltonians, and can be clearly seen in Fig. 2(a) for $N=10^4$.  This deflection shows that the Pincus regime is beginning to emerge, but the width of the regime $\D f$ is vanishingly small.}}  We also see the expected return to the non-interacting FJC behavior for large $f$.  The fit is, however, quite poor for $a\beta f\approx 1-4$, where the effective scaling exponent differs greatly from the simulation data.  
Figure 3(a) shows that the poor fit for intermediate $f$ originates with a slight overestimation in $\la Z\ra$ vs. f near $a\beta f\approx 1$, followed by an underestimation in $\la Z\ra$ near $a \beta f\approx3$.  This over- and underestimation produces a FEC that is not monotonically increasing with $f$, which is a completely non-physical result.  The small differences between the theoretical and simulated FECs are greatly exaggerated by the effective scaling exponent in the intermediate force range.  


The reason for the discrepancy between theory and simulation {{for intermediate forces}} is that, in the approximate representation for the (nearly) inextensible chain, extensions from $|\D\rv_n|=a$ are allowed {{(see Appendix B)}}.  As a result, the chain can stretch somewhat, with mean monomer spacing exceeding $a$.  For this reason, less force is required to extend the chain at intermediate forces, producing an overestimate of the FEC.  The minor disagreement between the theory and simulations in the FEC is amplified when the effective exponent $y=\partial \log(\,\la Z\ra\,)/\partial \log(\j)$ is computed (Fig 3(b)).  We see, however, that both the extensible and inextensible polymer models in a good solvent accurately predict the Pincus-like regime observed for $a\beta f\sim 10^{-2}-10^{-1}$.  At high forces, the response to the force depends on the precise model used to account for chain connectivity.  As a result, the predictions for the extensible and inextensible polymer models are vastly different when $a\beta f>1$.\\

\subsection*{Reexamination of the Blob Concept for Finite $N$ }

In order to better understand the unexpected scaling behavior of the FEC's for finite $N$, a more detailed study of the physical processes of extension are required.  There are three mechanisms by which the average extension of an extensible chain can increase as a function of force.  The first is orientation of the polymer along the force axis.  We expect that, for small $f$, the force will cause alignment with the $z$ axis, with little perturbation of the chain conformation.  In the second mechanism, the extension of the polymer is determined by an interplay between $\xi_P$ (a length scale below which $f$ is not relevant), and $N$ (which effectively determines the number of aligned blobs along the force direction).  We expect this mechanism will occur for intermediate forces, and for sufficiently large $N$, cause the emergence of the Pincus regime.  As these blobs are stretched, $\la Z\ra$ will increase without significantly affecting the alignment along the $f$ axis.  For large forces we expect overextension to dominate, when the chain is fully aligned and the monomers on a length scale $\xi_P$ are stretched.  In order to see these physical mechanisms of the extension in the simulations of finite, {{extensible}} polymers with $N=100$, we compute the effective force-induced alignment exponent $\omega$, given by $\la Z  / |\Rv|\ra\sim f^{\omega}$, and the effective overextension exponent $\mu$, given by $\la L/N\ra \sim f^{\mu}$.  If the polymer is perfectly aligned along the $z$-axis, we expect that the exponent $\omega \to 0$.  The variations of the effective exponents $\omega$ and $\mu$ for $N=100$ as $f$ changes are shown in Fig. 4(a).  We see that the polymer aligns with the $z$ axis at relatively small forces, with full alignment ($\omega \to 0$) occurring for $\j\approx 0.1$.  Overextension does not begin until $\j\approx 3$ (in the non-universal regime, see Fig 4(a)), giving a wide range of forces in which stretching of the monomers inside of the blobs contributes to the behavior of $\la Z\ra$.  Representative snapshots of the chain configuration in the three regimes are shown in Fig. 4(b).

The absence of a clear signature of the Pincus regime, even for $N=1600$, is intimately related to the breakdown of the inequality $N\gg(\xi_P/a)^{\frac{1}{\nu}}\gg 1$.  For large enough $N$, when the nonlinear regime in the FEC is observed (Fig 2(a)), the size of the blob $\xi_P\approx k_B T/f$ is expected to scale as $\xi_P\approx a N_b^{\nu}$, where $N_b$ (presumed to be much greater than unity) is the number of monomers inside of the blob.  The monomer density, $\rho_m$, inside the blob will scale as
\begin{equation} 
\rho_m \sim N_b/\xi_P^3 \sim \xi_P^{1/\nu-3} \sim (f/k_BT)^{3-1/\nu}.\label{blobscale}
\end{equation}
In good solvents, $\nu=3/5$, and hence $\rho_m$ is given by $\rho_m\sim f^{4/3}\equiv f^m.$  If the effective value for $m$ with finite $N$ exceeds $m=4/3$, as could be the case when the force locally stretches the chain segments inside $\xi_P$, we will find $\langle Z\rangle\sim f^x$ with $x\ne 2/3$ in the intermediate force regime. 

\begin{figure}
\begin{center}
\includegraphics[width=.6\textwidth]{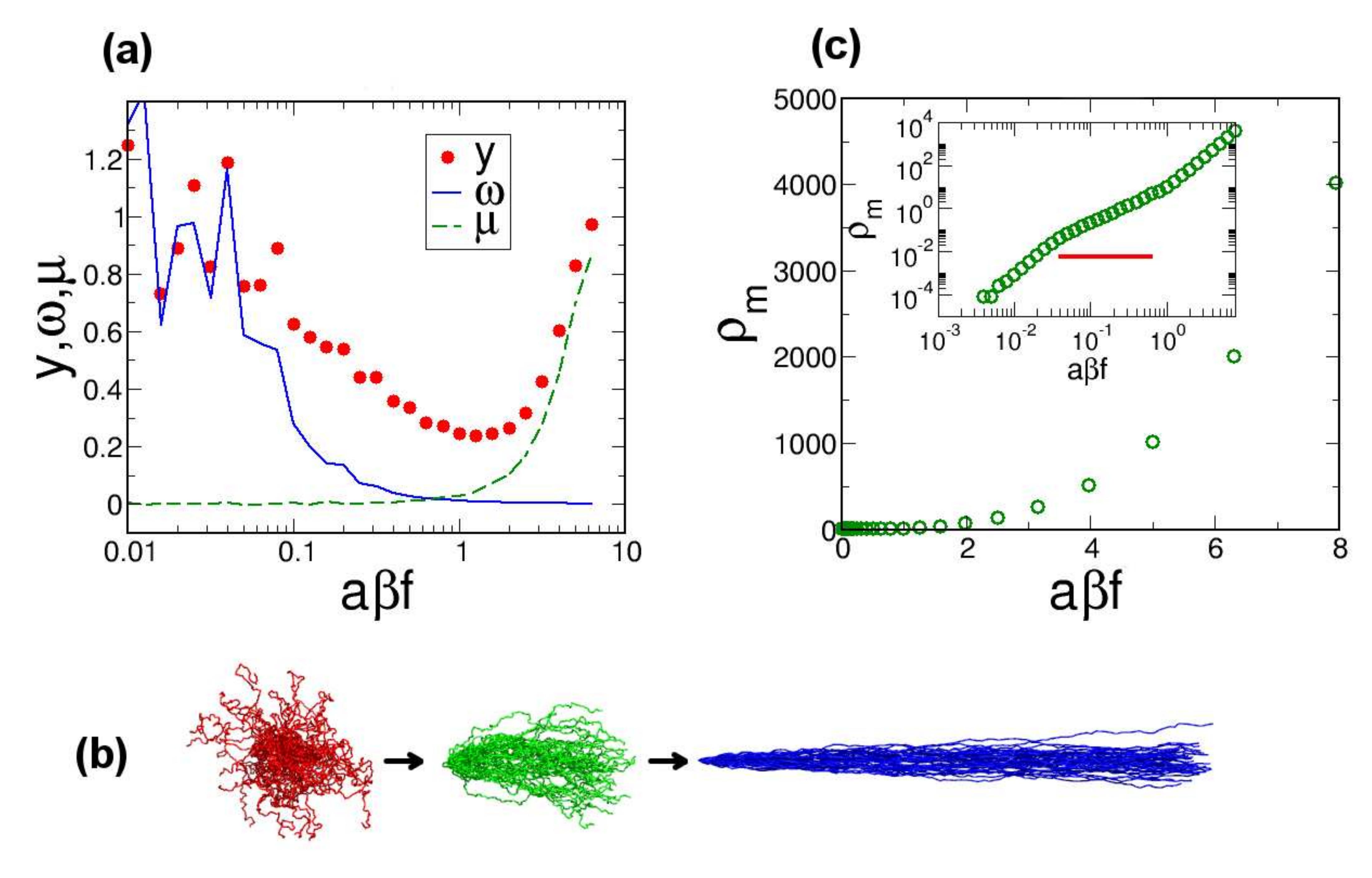}
\caption{(a)  Contributions to the effective scaling exponent $y$.  Shown are $y$ (dots), as well as the alignment exponent $\omega$ (solid line) and overstretching exponent $\mu$ (dashed line).  (b)  The blob density as a function of force. The inset shows a log-log plot of the monomer density inside the blob, showing three distinct scaling regimes.  {Scaling relation $\rho_m$ vs $f$ is obtained by fitting the data above red line in the inset.} (c)  The ensembles of structures at $a\beta f=0$ (red), 0.5 (green), and 8.0 (blue) are given to demonstrate the three step mechanism of the extensible chain stretching, i.e., (i) alignment, (ii) disruption of tensile blob, and (iii) overstretching. }
\end{center}
\end{figure}

In order to provide insights into the effective blob response to $f$ for the self-avoiding extensible polymer {of the finite size (N=100)}, we have calculated the dependence of the monomer density inside the blob on $f$.  To obtain the scaling behavior between monomer density and the force from the simulations, we perform the following steps:
\begin{enumerate}
\item Make a sphere of radius $b=\xi_P/2$, with $\xi_P(=k_BT/f)$, {{centered on}} the $i^{th}$ monomer and count the number of monomers ($N_b$) within the sphere whose volume is $b^3$. The density of monomers within the sphere center at $i^{th}$ monomer is $\rho_m(i)\sim N_b/b^3$. 
\item Move to the $(i+1)^{th}$ monomer, and compute the density again. 
\item When $i=N$, the average density is computed using $\langle\rho_m(f)\rangle=1/N\sum_{i=1}^N\rho_m(i)$.
\item Repeat this procedure for the ensemble of structures obtained at each force. 
\end{enumerate}
Although this method of computing the monomer density from the polymer structures 
is very crude, the scaling exponent between $\rho_m$ and $f$ should not be affected by the details of the calculation.  The results are shown in Fig. 4(c).  We find that $\rho_m \sim f^{1.6}$ in the intermediate force regime (data above the red base line in the inset of Fig. 4(c)) .   From Eq. (\ref{blobscale}), a density scaling of $f^{1.6}$ implies $\xi_P\sim N_b^{0.71}\ne N_b^{0.60}$, which indicates that there is no force range in which ideal blobs can be observed for small $N$.  In other words, the separation in length scale $N\gg N_b\gg1$ is not satisfied.  The observed scaling exponent for $N_b$ is greater than that for a simple self-avoiding walk, which suggests that the monomers inside of the blob do not behave as unperturbed SAW's.  Thus, the fundamental premise used in the blob argument used to derive the Pincus regime breaks down for small $N$.  The tensile force is felt by the monomers within the blobs, which swell due to the stretching of monomers inside $\xi_P$.  The density of monomers inside the blob scales differently than the expected for large values of $N$, and provides the microscopic reason why, in the finite-sized self-avoiding chain, $\langle Z \rangle\sim f^x$ with $x<2/3$.  As $N$ increases, the intermediate force regime can be large enough so that $\rho_m\sim f^{\frac{4}{3}}$, which is needed to see the Pincus scaling $\la Z\ra\sim f^{\frac{2}{3}}$.\\

\section*{III. Homopolymer in a Poor Solvent}

\subsection*{Theory}
In a poor solvent, the second virial coefficient ($v_0$) becomes negative.  The strength of the attractive interactions between the monomers exceed that between the monomers and the solvent.  As a result, the polymer adopts collapsed, globular conformations at temperatures below the Flory $\Theta$ temperature.  In poor solvents, the Edwards model is modified to include an effective three-body interaction, to ensure that the averages of physical observables {{converge}}.  The {{extensible}} Hamiltonian {{in a poor solvent}} is $\beta H_P=\beta H_0+\D_3$, where $H_0$ is defined in Eq. ($\ref{HG}$) and 
\begin{eqnarray}
\D_3=\frac{w_3}{6}\int_0^N ds\int_0^N ds'\int_0^N ds'' \ \d[\rv(s)-\rv(s')]\ \d[\rv(s')-\rv(s'')].
\end{eqnarray}
The self-consistent equation for the extension {{in this case}} becomes $\la Z(\D_1+\D_2+\D_3)\ra=\la Z\ra\la\D_1+\D_2+\D_3\ra$, {{similar to Eq. (\ref{sc})}}.  We have already determined the $\D_1$ and $\D_2$ terms, and need only compute $\la Z\D_3\ra-\la Z\ra\la\D_3\ra=a \partial/\partial \j\la \D_3\ra$.
The SCE for {{an extensible polymer in a poor solvent}} can be  written as,
\begin{eqnarray}
\l^2-1&=&\frac{v\sqrt{N}}{\l^3}\int_\d^1 du\ \frac{1-u}{\sqrt{u}}e^{-N\l^2\j^2 u/6}\nonumber\\
&&\qquad\qquad+\frac{w}{\l^6}\int_\d^1 du_1 \int_\d^{1-u_1}du_2\frac{(1-u_1-u_2)(u_1+u_2)}{u_1^{3/2}u_2^{3/2}}e^{-N\l^2\j^2(u_1+u_2)/6},\label{scpoor}
\end{eqnarray}
where we have defined $w=(3/2\pi)^3\ w_3/a^6$.  {{The inextensible self-consistent equation is similar, and has a similar root structure, and we will therefore omit such a calculation here}}.  Again, we have included a cutoff in the integrals, as was done for the two-body case.  However, the three-body integral in Eq. (\ref{scpoor}) is clearly divergent for $\d=0$, unlike the two-body term.  {{This divergence must be removed for the self-consistent equation to converge in the limit of $N\to\infty$, by renormalizing $w$.  For $f=0$, we can evaluate the three-body integral exactly, and find that, with $\d\sim \l/N$ and as $N\to\infty$, it diverges as $\sim 16/3\sqrt{\d}$.  The $N\to\infty$ divergence is therefore removed if we renormalize $w=\bar w/\sqrt{N}$.}}  It is not difficult to show that the self-consistent equation has a solution $\l\approx \l_{g}=(4\bar w/|v|)^{\frac{1}{3}}N^{-\frac{1}{6}}$ for $f=0$ and large $N$, giving the expected scaling $\la\Rv^2\ra\sim N^{\frac{2}{3}}$ for a homopolymer in a poor solvent.  However, the final term of Eq. (\ref{scpoor}) can not be evaluated exactly for non-zero $\j$, so we must resort to numerical work in order to determine the roots for larger forces.  

\begin{figure}
\begin{center}
\includegraphics[width=0.9\textwidth]{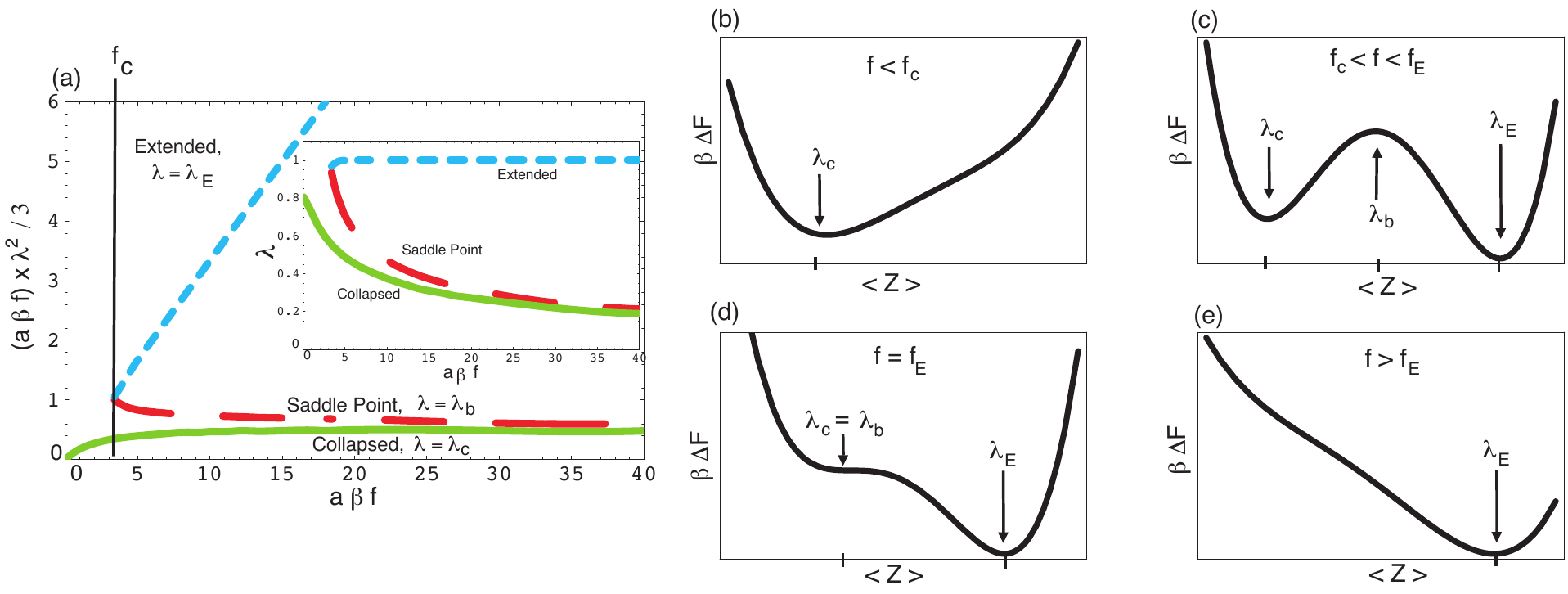}
\caption{(a):  The extension $\la Z\ra/L=(a\beta f)\l^2/3$ for the three roots of the self-consistent equation in a poor solvent for $v=-5$ and $\bar w=1$, with $\delta=1.6\l/N$:  $\l_g$ (solid line), $\l_b$ (dashed line), and $\l_E$ (dotted line).  The three values for $\l$ are shown in the inset.  (b)  For $f<f_c$, the polymer is globular.  (c) In the force range $f_c<f<f_E$, the chain conformations are a combination of globular and extended states.  (d)  at $f=f_E$, the globular configuration is marginally stable.  (e)  For $f>f_E$, the chain is in the fully extended state.}
\end{center}
\end{figure}

We find that Eq. (\ref{scpoor}) has three unique roots beyond a critical force $f_{c}${{, which correspond to collapsed ($\l_c$), extended ($\l_E$), and saddle point ($\l_b$) structures.}}  Numerically, we find $0<\l_{c}\le \l_{g}$, and $\l_{E}\approx 1$ for $f>f_c$.  Our interpretation of $\l_c$ as corresponding to a collapsed state is only qualitative, because an extensible homopolymer (used as the reference Hamiltonian in the calculations) in a poor solvent does not have a unique `collapsed' state.   With the interpretation that $\l_{c}$ and $\l_{E}$ are the roots signifying the two local minima of the free energy for the collapsed and extended states, we can interpret the saddle point solution $\l_b$ as a local maximum in the free energy, i.e. the barrier (or saddle point) between the two states.  Again, this interpretation is qualitative only, because there is no well defined `barrier' between the collapsed and extended states.  In Fig. 5(a) we show the extension $\la Z\ra/L= \j\l^2/3$ for the three solutions to the self-consistent equation (\ref{scpoor}) for $v=-5$ and $\bar w=1$ (arbitrarily chosen), and with $\d=1.6\l/N$.  We see $a\beta f_c\approx 3.5$ is the critical force at which the extended and saddle point solutions emerge.  The critical force $f_c$ depends on the particular values of $v$ and $\bar w$, and we expect it will be an increasing function of $|v|/\bar w$.  In this triple-root regime, the polymer will be in bistable equilibrium between the collapsed ensemble and extended state, suggesting the development of a pearl-necklace structure for intermediate $f$.  The collapsed and saddle point solutions coalesce for a finite $f=f_E$ (Fig 5(a) inset).  For $f>f_E$, $\l_c$ and $\l_b$ vanish, leaving the extended root $\l_E$ the only solution to Eq. (\ref{scpoor}).  This shows, as expected, that the inter-monomer interactions become irrelevant for sufficiently high force, and $\la Z\ra\sim Na\times(a\beta f)/3$ as $f\to\infty$.  Schematic pictures of the free energy as a function of the extension $\la Z\ra$ (Fig. 5(b-e)) for varying force illustrate our qualitative interpretation of the solutions to the self-consistent equation (\ref{scpoor}).

A similar multi-root structure has been previously predicted for a polymer in a poor solvent with electrostatic interactions \cite{HaThirum,DuaVig,Scheissel}.  These references note the emergence of multiple roots beyond a critical value of the backbone charge density (in this respect, equivalent to the tension), and qualitatively identify the meaning of the multiple roots as we have.  However, because the Edwards Singh method can not predict the barrier height or the depth of the minima, we can not quantitatively predict $\la Z\ra$ for a polymer in a poor solvent.  The qualitative picture, namely the tension-induced globule to rod transition which should occur when $f>f_c$, is confirmed using explicit simulations of force-induced stretching of a homopolymer in a poor solvent.  The simulations (see below) also provide a microscopic picture of the structural transitions that occur as $w_3$, in Eq. (\ref{scpoor}), increases.\\

\subsection*{Simulations }
The simulation procedure used to study the stretching of a homopolymer in a poor solvent is identical to the one described for the good solvent case, except for the Hamiltonian used. 
The Hamiltonian in a poor solvent is 
\begin{equation}
\beta H=\frac{3}{2a^2}\sum_{i=1}^{N-1}(|\rv_{i+1}-\rv_i|^2-a^2)+\sum_{i=1}^{N-2}\sum_{j=i+2}^N\varepsilon \left[\left(\frac{a}{|\rv_i-\rv_j|}\right)^{12}-2\left(\frac{a}{|\rv_i-\rv_j|}\right)^6\right]-\beta f(z_N-z_1),
\label{eqn:Hamiltonian_poor}
\end{equation}
where $\varepsilon=0.5$ and $1.5$ are used for different solvent conditions, and where the other parameters are the same as in the good solvent case.  
The nature of the polymer is characterized by the second virial coefficient $v_2=\int d\textbf{r}\{1-e^{-\beta V_{int}(r)}\}$, where $V_{int}(r)$ is the second term of Eq. (\ref{eqn:Hamiltonian_poor}).
When $\varepsilon\approx 0.3$, $v_2$ approaches zero, and corresponds to the theta condition ($T=T_{\Theta}$). 
By decomposing $v_{int}(r)$ into repulsive ($v_{rep}(r)$) and attractive ($v_{rep}(r)$) parts of the potential, one can write $v_2\approx \int d{\bf r}\{1-e^{-\beta v_{rep}(r)}\left(1-\beta v_{att}(r)\right)\}=v_0\left(1-T_{\Theta}/T\right)$ where $v_0=\int d{\bf r}\left(1-e^{-\beta v_{rep}(r)}\right)$ \cite{EdwardsBook}. Therefore, 
\begin{equation}
T_{\Theta}\approx T\left(1-\frac{v_2}{v_0}\right). 
\end{equation}
We find $v_2=-1.9a^3$ ($T_{\Theta}\approx 1.7\times T$, weakly-hydrophobic condition) for $\varepsilon=0.5$ and $v_2=-15.2a^3$ ($T_{\Theta}\approx 6.4\times T$, hydrophobic condition) for $\varepsilon=1.5$.  These estimates for $T_\Theta$ as a function of $\varepsilon$ are approximate.  For our purposes, approximate estimates are sufficient to illustrate the response of weakly hydrophobic and strongly hydrophobic chains to force.

\begin{figure}
\begin{center}
\includegraphics[width=.7\textwidth]{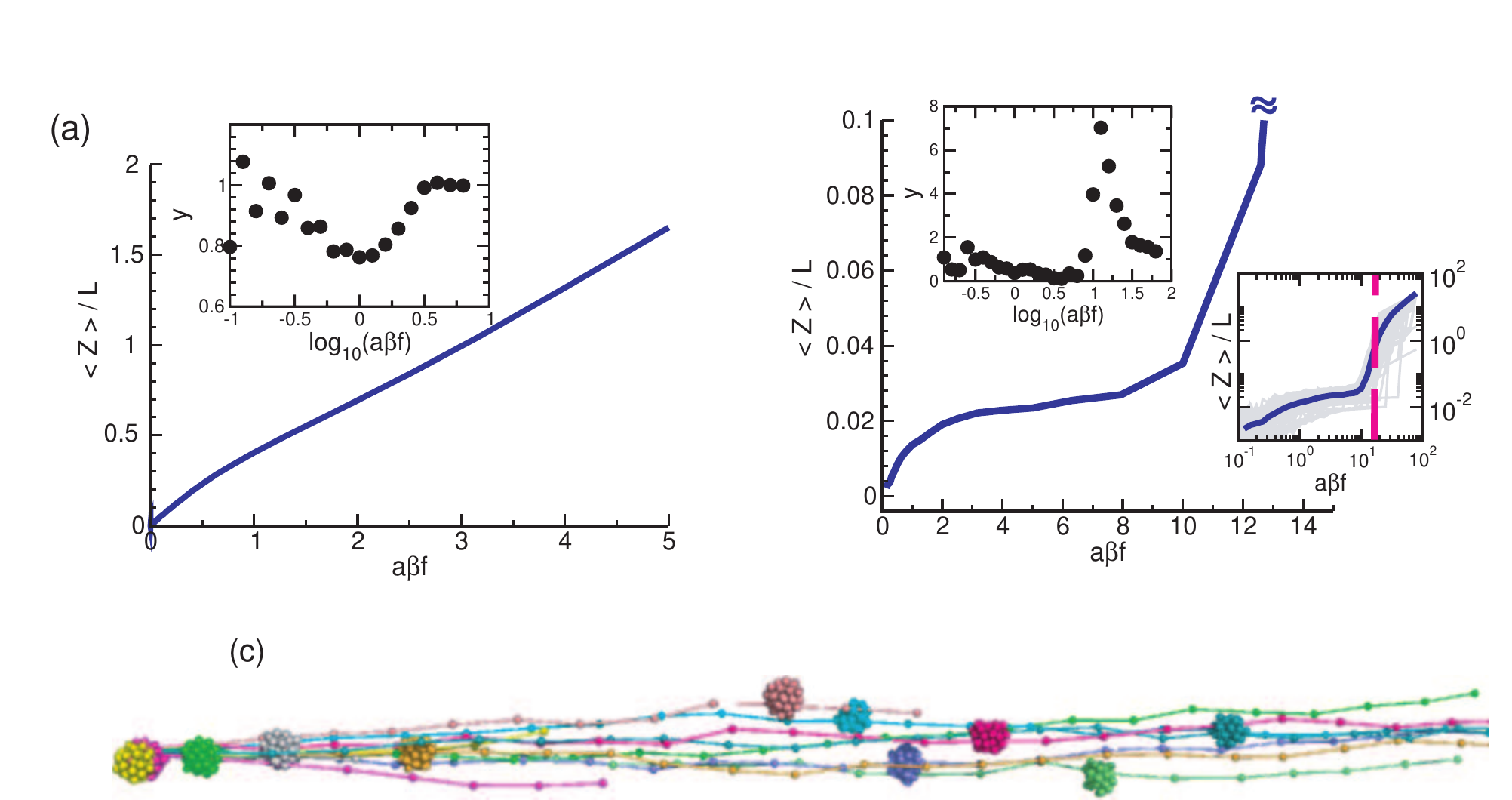}
\caption{(a)  Extension as a function of force for a weakly hydrophobic polymer ($\varepsilon$ in Eq. $(\ref{eqn:Hamiltonian_poor}$) is 0.5).  (b)  Same as (a), except the chain is strongly hydrophobic ($\varepsilon=1.5$).  The insets show the effective scaling exponent $y$ (Eq. ($\ref{ExpDef}$)).  The transition to the extended state in (a) appears continuous.  For the strongly hydrophobic polymer, the globule $\to$  rod like transition is sharp.  The transition force depends on the energetic details of the globule.  The heterogeneity of the transition is manifested as the broad variations of transition force. The ensemble of structures found at the globule-to-rod transition force ($f_c=1.8k_BT/a$) are shown in (c).}
\end{center}
\end{figure}

In Figure 6, we show the average linear extension for weakly hydrophobic (a) and strongly hydrophobic (b) polymers as a function of force.  The weakly hydrophobic polymer does show a transition between two linear scaling regimes, with the low force behavior of $\la Z\ra\approx \la\Rv^2\ra_{f=0}\times (\beta f)/3$, and the high force behavior returning to the non-interacting $\la Z\ra=N a^2\beta f/3$.  The transition is very smooth, and does not show the expected first order transition due to the weak nature of the interactions, as shown in the inset.  The strongly hydrophobic chain does show a first order transition around $a \beta f_c\approx 1.8$, {but with broad dispersion. Variations in the critical unbinding force is substantial from molecule to molecule, due to the microscopic heterogeneity of the globular structures.} 
{{The observed plateau in Fig 6(b) is most likely due to full alignment of the globule along the $z$-axis, as was the case for the self-avoiding polymer (Fig 4(a)), and seen in the theoretical predictions (Fig 5).  There is a large range of forces over which the FEC does not resemble either the globular or fully extended states, showing the bistable equilibrium between the two.}} \\

\section*{VI. The Scattering Function Under Force}
{The analysis using scattering experiments is useful for investigating the overall polymer configurations, because the scattering intensity as a function of momentum transfer ($I(q)=\la 1/N^2$ $\sum_{i<j}\exp{\left(i\mathbf{q}\cdot \rv_{ij}\right)\ra}$) provides structural information on all length scales. In contrast, the FEC only provides information about the extension of the chain. By comparing with the well-known scaling relations of $I(q)$ with respect to $q$ for various shapes, one can obtain the structures of the polymer over all length scales. For example, for the various structures we expect $I(q)\sim q^{-x}$, with $x=2$ (Gaussian chain), $x=1$ (rod), $x=4$ (globules), and $x=5/3$ (polymer in a good solvent) \cite{HigginsBook}.
We calculated the scattering intensity by integrating the distance distribution function obtained from the ensemble of structures,  
\begin{eqnarray}
I(q,f)=\int d^3{\rv}P({\rv},f)e^{i{\qv}\cdot {\rv}}=4\pi\int_0^{\infty}dr\  r^2P(r,f)\ \frac{\sin{qr}}{qr},
\end{eqnarray} 
with $q=|\qv|$.  In our simulations performed under varying tension values, we obtained $4\pi r^2 P(r,f)$ directly from the ensemble of structures by collecting the histograms between the interval of $(r,r+dr)$ with $dr=0.2 a$.}
 
 \begin{figure}
\begin{center}
\includegraphics[width=1\textwidth]{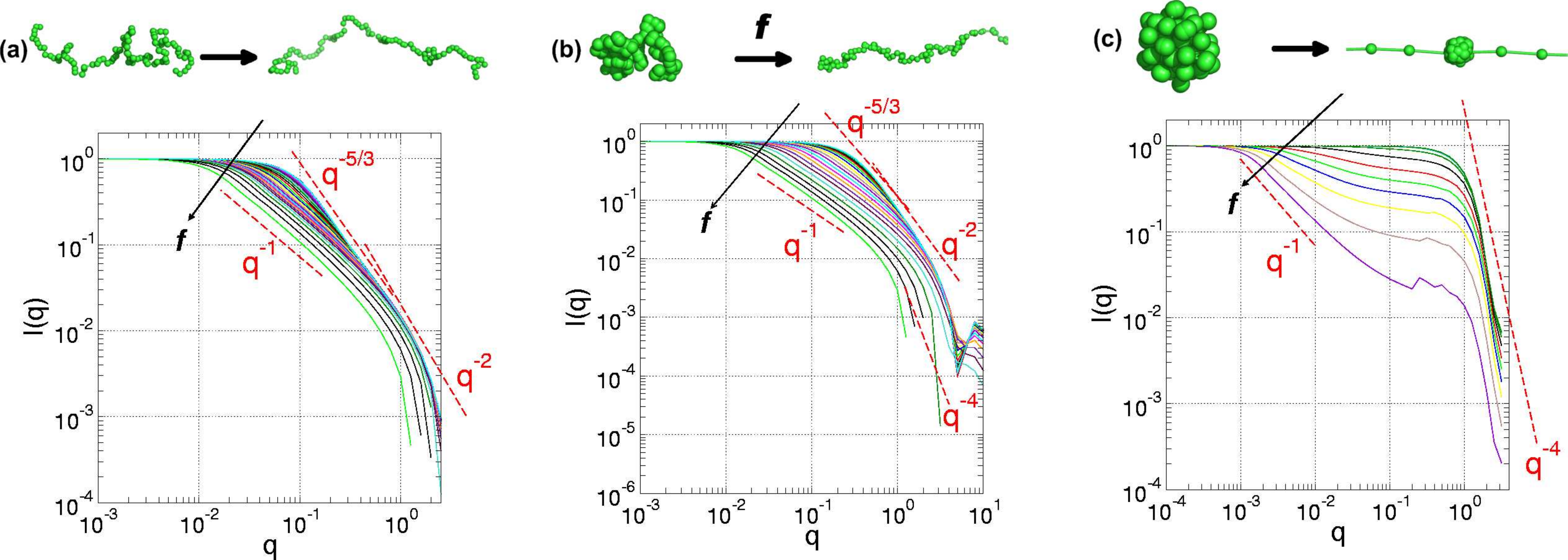}
\caption{(a) $I(q)$ for a homopolymer in good solvent under varying tension, (b) $I(q)$ for weakly hydrophobic homopolymer under tension, (c) $I(q)$ for strongly hydrophobic homopolymer under tension.  The arrows in (a), (b), and (c) indicate increasing $f$ values.
The tension-induced structural changes of a homopolymer are illustrated in three solvent conditions (good, near theta, and poor solvent conditions).}
\end{center}
\end{figure}

{An inspection of the scattering intensity $I(q)$ of a homopolymer in different solvent conditions, {{shown in Fig. 7, along with snapshots of representative structures}}, succinctly summarizes the shapes adopted by the polymer as a result of the tension induced structural transitions.  
(i) In good solvents {{(Fig 7(a))}}, the entire chain of N=100 is characterized by the tensile blob in the absence of force (or for small force), with $I(q)\sim q^{-5/3}$ for $q\sim 0.1-1$.  As $f$ increases, the tensile blobs continuously change to the rod state, which is indicated by $I(q)\sim q^{-1}$. 
(ii) For the weakly hydrophobic condition {{(Fig 7(b))}}, i.e., slightly above the theta temperature, the chain displays a hierarchy of structures on distinct length scales.  When $f$ is small, both signatures of Gaussian coil ($I(q)\sim q^{-2}$) and globule structure ($I(q)\sim q^{-4}$) are found on small length scales $q^{-1}\lesssim 1$, while the chain is characterized by the polymer in a good solvent for $q^{-1}\gtrsim 1$.  As $f$ increases, the globule to rod transition of the self-avoiding chain takes place continuously. 
(iii) For the strongly hydrophobic condition {{(Fig 7(c))}}, the whole chain is collapsed to compact globule ($I(q)\sim q^{-4}$. The globular structure is maintained so that all $I(q)$'s are practically identical for $q^{-1}\gtrsim 1$ as long as $f<f_c$. When $f$ becomes greater than $f_c$, a sharp transition occurs, reflecting the globule ($I(q)\sim q^{-4}$) to the rod ($I(q)\sim q^{-1}$) transition.}  The first order nature of force-induced stretching has been previously described using scaling arguments \cite{Halperin}.
\\ 

\section*{V. Conclusions}

We have developed a general theory for describing the response of homopolymers to {{an external}} force for arbitrary values of $N$, the number of monomers.  By using both an extensible and inextensible model for the polymer in a good solvent, we show that the theoretical results are in accord with the predictions of the Pincus scaling laws.  The mean chain extension depends linearly on the force for small $f$, and scales as $\la Z\ra\sim f^{\frac{2}{3}}$ for intermediate $f$ and {{sufficiently large}} $N$.  Simulations of an extensible chain with $N=100$ and the thick chain model with $N=1600$ were performed to validate the theory.  The theoretical predictions for the force-extension curves are in excellent agreement with the simulation results.  Surprisingly, the expected Pincus scaling is not observed in simulations, even for $N=1600$.  The theory predicts that the width $\D f$ for observing the Pincus regime for $N\sim O(10^3)$ is vanishingly small.  Only when $N$ exceeds $
\sim10^5$ can the strong stretching limit ($\la Z\ra\sim f^{\frac{2}{3}}$) be unambiguously observed.  The failure to observe the Pincus scaling is linked to the breakdown of the notion that the monomers inside the well-defined tensile blobs are unperturbed.  For $N\sim O(10^3)$, the monomers inside each blob feel the effect of force, which essentially violates the required inequality $N\gg(\xi_P/a)^{\frac{1}{\nu}}\gg 1$.  

{{Applying tension to a polymer in a poor solvent produces a much richer set of structures,}} because of the presence of an additional attractive monomer-monomer energy scale.  In the absence of force, a polymer in a poor solvent forms a globule at $T<T_\Theta$.  For this case, the theoretical analysis predicts that the globule to stretched (i.e. rod-like conformation) transition should occur abruptly via a first-order transition when $f$ exceeds a critical force.  While the simulation results are in accord with the theoretical predictions, they show several structural transformations, depending on the quality of the solvent.  The hierarchy of structures are reflected in the force-dependent structure factor.  For weakly hydrophobic polymers ($T\approx T_\Theta^+$) for small forces, the scattering function $I(q)$ shows signatures of Gaussian and globular structures at large length scales (small $q$), whereas over small length scales the polymer behaves as a self-avoiding chain.  At large forces, the transition to a rod-like conformation occurs.  These structural transitions occur continuously as $f$ increases  for a weakly hydrophobic chain.  Strongly hydrophobic chains ($T<T_\Theta$) adopt globular structures for small forces.  The conformation remains globular as long as $f<f_c\approx k_B T_\Theta/R_g$.  The globular nature of the conformation is reflected in the $I(q)\sim q^{-4}$ scaling.  If $f>f_c$, there is an abrupt transition to the rod-like state, which is reflected in the $I(q)\sim q^{-1}$ scaling. 

The predictions made here can be, in principle, validated with single molecule AFM or optical tweezers experiments.  Our simulations show that the forces required to stretch the homopolymer ($N\approx 100$) is on the order of about 30 pN, which are easily accessible in current experiments.  
\\

\section*{VI. Appendix A: Self-Consistent equation for $\l$}
{
In this appendix, we provide the details for the calculations of $\la Z\ra_1$ and $\la\delta[\rv(s)-\rv(s')]\ra_1$ that appear in Eqs. (\ref{Del1}) and (\ref{Del2}). 
\begin{align}
\la Z\ra_1&=\frac{\int\mathcal{D}\rv(s)Ze^{-\beta H_1}}{\int \mathcal{D}\rv(s) e^{-\beta H_1}}=\frac{\partial}{\partial(\beta f)}\log\bigg[{\int\mathcal{D}\rv(s)e^{-\beta H_1}}\bigg]\nonumber\\
&=\frac{\partial}{\partial(\beta f)}\log\bigg[{\int\mathcal{D}\rv_{\perp}(s)
e^{-\frac{3}{2a^2\l^2}\int_0^Nds\rv_{\perp}^2(s)}\int\mathcal{D}z(s)
e^{-\frac{3}{2a^2\l^2}\int^N_0ds\left(\dot{z}(s)-\frac{a^2\l^2\beta f}{3}\right)^2+\frac{Na^2\l^2}{6}\beta^2 f^2}}\bigg]
\nonumber\\
&=\frac{1}{3}Na^2\l^2\beta f
\end{align}
\begin{align}
\la\delta[\rv(s)-\rv(s')]\ra_1 &=\int\int\int\int d\rv(N)d\rv(s')d\rv(s)d\rv(0)\ G(\rv(N)-\rv(s')|N-s')\delta[\rv(s)-\rv(s')]\nonumber\\
&\qquad\qquad\qquad\qquad\times G(\rv(s')-\rv(s)|s'-s)G(\rv(s)-\rv(0)|0)\nonumber\\
& =G(0|s'-s),
\end{align}
where the propagator $G(\cdots)$ is decomposed into transverse and longitudinal components, $G(\Rv|N)=G_{\perp}(\Rv_{\perp}|N)G_{\parallel}(Z|N;f)$, each of which can be exactly obtained for the reference Hamiltonian.  We find 
\begin{eqnarray}
G_{\perp}(\Rv_{\perp}|N)&=&\frac{\int\mathcal{D}\rv_{\perp}(s)\delta\left(\int^N_0ds\dot{\rv}_{\perp}(s)-\Rv_{\perp}\right)e^{-\frac{3}{2a^2\l^2}\int^N_0ds\dot{\rv}^2(s)}}{\int\mathcal{D}\rv_{\perp}(s)e^{-\frac{3}{2a^2\l^2}\int^N_0ds\dot{\rv}^2(s)}}\nonumber\\
&=&\frac{\int\mathcal{D}\rv_{\perp}(s)\int\frac{d^2\mathbf{k}}{(2\pi)^2}e^{i\mathbf{k}\cdot\left(\int^N_0ds\dot{\rv}_{\perp}(s)-\Rv_{\perp}\right)}e^{-\frac{3}{2a^2\l^2}\int^N_0ds\dot{\rv}^2(s)}}{\int\mathcal{D}\rv_{\perp}(s)e^{-\frac{3}{2a^2\l^2}\int^N_0ds\dot{\rv}^2(s)}}\nonumber\\
&=&\int\frac{d^2\mathbf{k}}{(2\pi)^2}e^{-\frac{Na^2\l^2}{6}\left(\mathbf{k}-\frac{3\Rv_{\perp}}{Na^2\l^2}\right)^2-\frac{3\Rv_{\perp}^2}{2Na^2\l^2}}\nonumber\\
&=&\left(\frac{3}{2\pi Na^2\l^2}\right)\exp{\left(-\frac{3\Rv_{\perp}^2}{2Na^2\l^2}\right),}
\end{eqnarray}
where the Fourier representation of the $\delta$-function used, and then the path integral is evaluated. The propagator $G_{\perp}(\Rv_{\perp}|N)$ is obtained after performing the integral in Fourier space.  
The longitudinal component is also similarly calculated for the Hamiltonian with linear force term, 
\begin{eqnarray}
G_{\parallel}(Z|N;f)&=&\frac{\int\mathcal{D}z(s)\delta\left(\int^N_0ds\dot{z}(s)-Z\right)e^{-\frac{3}{2a^2\l^2}\int^N_0ds\dot{z}^2(s)+\beta f\int^N_0ds\dot{z}(s)}}{\int\mathcal{D}z(s)e^{-\frac{3}{2a^2\l^2}\int^N_0ds\dot{z}^2(s)+\beta f\int^N_0ds\dot{z}(s)}}\nonumber\\
 &=&\left(\frac{3}{2\pi Na^2\l^2}\right)^{1/2}\exp{\left[-\frac{3}{2Na^2\l^2}\left(Z-\frac{Na^2\l^2\beta f}{3}\right)^2\right]}.
\end{eqnarray}
Thus, we find
\begin{equation}
\la\delta[\rv(s)-\rv(s')]\ra_1=G(0|s'-s)=\left(\frac{3}{2\pi |s'-s|a^2\l^2}\right)^{3/2}\exp{\left[-\frac{|s'-s|a^2\l^2\beta^2 f^2}{6}\right]}.
\end{equation} 
}

In obtaining Eq. ($\ref{sc}$), we used the extension {$Z$} as the observable to determine the optimal value of $\l$.  
{Alternatively, one can also obtain the SCE for $\l$ using the transverse fluctuation of polymer $\Rv_{\perp}^2=X^2+Y^2$  where $X$ and $Y$ are the projections of the end-to-end distance vector $\Rv$, i.e., $\la \Rv_\perp^2 (\D_1+\D_2)\ra-\la\Rv_\perp^2\ra\la\D_1+\D_2\ra=0$.}  Computations involving $\Rv_\perp^2$ are significantly simpler than those involving the end-to-end distance vector, $\Rv^2$, because the propagators in the $x$ and $y$ directions are decoupled from the force-dependent propagator in the $z$ direction.  Using the same methods as before with our original variational Hamiltonian in Eq. (\ref{HV}), we find 
\begin{eqnarray}
\la \Rv_{\perp}^2\D_1\ra-\la\Rv_{\perp}^2\ra\la\D_1\ra&=&\frac{2Na^2\l^2(\l^2-1)}{3},\label{RvDel1}\\
\la \Rv_{\perp}^2\D_2\ra-\la\Rv_{\perp}^2\ra\la\D_2\ra&=&-\frac{a^2\l^2v_0}{3}\int^N_0ds\int_0^Nds'\,G(0|s'-s)(s'-s).\label{RvDel2}
\end{eqnarray}
Using Eqs. (\ref{RvDel1}) and (\ref{RvDel2}) we obtain the SCE for $\l$,
\begin{eqnarray}
\l^2-1=\frac{v\sqrt{N}}{\l^3}\int_\d^1 du\frac{1-u}{\sqrt{u}}\ e^{-Nu\l^2\j^2/6 }\ ,
\end{eqnarray}
which is identical to the equation obtained using the linear end-to-end distance {($Z$)} {{as the generating observable}} in Eq. (\ref{GaussianConsistent}).  Thus, the computation of the FEC is not dependent on whether $Z$ or $\Rv^2_\perp$ is used in determining the self-consistent equation.
\\

\section*{VII. Appendix B:  The Thick Chain Model}

In order to verify the theoretical predictions for the polymer described by the (nearly) Inextensible Gaussian Hamiltonian (IGH) with excluded volume interactions, we have simulated the FEC using the thick chain (TC) model for a self-avoiding polymer.  In the TC model, the polymer is viewed as a chain with a finite uniform thickness $D$, and is represented as a succession of beads with position vectors ${\rv_0,..., \rv_N}$. All of the bond vectors $\Delta \rv_n = \rv_{n+1} - \rv_n$ ($n = 0,..,N$) have the same modulus $a$. Therefore, unlike the IGH where $\la|\D\rv_n|\ra\approx a$ in Eq. (\ref{InextensH}), {{the bond length restriction $|\D\rv_n|=a$}} is strictly enforced in the TC model.  The interaction potential of the TC under tension is given by
\begin{eqnarray}\label{eq:HamilTC}
  \mathcal{H}_{TC} &=& \sum_{i,j,k} V(R_{i,j,k}) -  \fv\cdot (\rv_N-\rv_0),
\end{eqnarray}
where the first term enforces the self-avoidance, and the second term represents the external force. In particular,
\begin{eqnarray}\label{eq:V3TC}
V(R_{ijk}) =
\left\{
  \begin{array}{ll}
    0, & \hbox{$R_{ijk} > D$} \\
    \infty, & \hbox{$R_{ijk} \le D$,}
  \end{array}
\right.
\end{eqnarray}
where $R_{ijk}$ is radius of the circle going through the triplet of beads $(i,j,k )$. Physically, the first term in the Hamiltonian (Eq. (\ref{eq:HamilTC})) ensures the self-avoidance of the chain by rejecting both local self-intersection (the local radius of curvature must be no smaller than $D$) and interpenetration of any two portions of the chain at some finite arc-length. Intuitively, it allows only configurations satisfying the thickness constraints, that the radii of circles going through all the triplets of beads $(i,j,k)$ be greater than $D$.

In order to characterize the stretching response of a thick chain with $D/a = 1$ and $N = 1600$ we performed Monte Carlo simulations using the following scheme.  Starting from an arbitrary initial
chain conformation that satisfies the thickness constraints, the exploration of the available configuration space was performed by distorting conformations by means of pivot and crankshaft moves. The new structures were accepted or rejected according to the standard Metropolis criterion (the infinite strength of the three-body penalties of Eq. (\ref{eq:HamilTC}) was enforced by always rejecting configurations violating the circumradius constraints). The relative elongation of the chain was calculated for increasing values of the applied stretching force. For each run, after equilibration, we measure the autocorrelation time and sampled a sufficient number of independent conformations to achieve a relative error of at most $10^{-3}$ in the average chain elongation. For moderate to high forces, this typically entailed the collection of $10^4$ independent structures, whereas a 10-fold increase of sampling was required for small forces due to the broad distribution of the end-to-end separation along the force direction.  For small forces, conformational fluctuations can be even larger than the mean extension, which makes achieving converged results for $\la Z\ra$ more difficult.

\section*{Acknowledgments}

This work was supported in part by a grant from the National Science Foundation through grant number CHE 05-14056.   

\clearpage

\end{document}